\documentclass[runningheads]{llncs}

\usepackage{graphicx}
\usepackage{cite}
\usepackage[misc]{ifsym}
\usepackage{hyperref}

\usepackage[ruled,vlined]{algorithm2e}
\include{pythonlisting}

\usepackage{amssymb}
\usepackage{amsmath}
\usepackage{array}
\usepackage{pifont}
\newcommand{\cmark}{\ding{51}}%
\newcommand{\xmark}{\ding{55}}%
\usepackage{multirow}
\usepackage{booktabs,makecell}
    
\usepackage{siunitx}

\usepackage[labelfont=bf]{caption}

\usepackage[labelformat=simple]{subcaption}

\begin{document}
%
\title{Measuring Financial Time Series Similarity With a View to Identifying Profitable Stock Market Opportunities}

\titlerunning{Financial Time Series Similarity}
%

 \author{Rian Dolphin\inst{1}\textsuperscript{(\Letter)} \and
 Barry Smyth\inst{1,2} \and
 Yang Xu\inst{3} \and
 Ruihai Dong\inst{1,2}}
%
\authorrunning{R. Dolphin et al.}
%
\institute{School of Computer Science, University College Dublin, Dublin, Ireland\\
\email{\href{mailto:rian.dolphin@ucdconnect.ie}{rian.dolphin@ucdconnect.ie}} \and
Insight Centre for Data Analytics, University College Dublin, Dublin, Ireland\\
\email{\{barry.smyth, ruihai.dong\}@ucd.ie} \and
School of Economics and Management, Beihang University, Beijing, China\\
\email{yang\_xu@buaa.edu.cn}}


%
\maketitle              

\begin{abstract}
Forecasting stock returns is a challenging problem due to the highly stochastic nature of the market and the vast array of factors and events that can influence trading volume and prices. Nevertheless it has proven to be an attractive target for machine learning research because of the potential for even modest levels of prediction accuracy to deliver significant benefits. In this paper, we describe a case-based reasoning approach to predicting stock market returns using only historical pricing data. We argue that one of the impediments for case-based stock prediction has been the lack of a suitable similarity metric when it comes to identifying similar pricing histories as the basis for a future prediction --- traditional Euclidean and correlation based approaches are not effective for a variety of reasons --- and in this regard, a key contribution of this work is the development of a novel similarity metric for comparing historical pricing data. We demonstrate the benefits of this metric and the case-based approach in a real-world application in comparison to a variety of conventional benchmarks.


\end{abstract}

\section{Introduction}
The stock market represents a challenging target when it comes to analysis and prediction~\cite{bachelier1900theorie,fama1970efficient,fama1965behavior}. The stochastic nature of stock prices reflects a complex network of interactions involving a web of hidden factors and unpredictable events. At the same time, the potential to identify even fleeting patterns in market data promises tremendous rewards and in a world where nanoseconds count even a modest degree of prediction accuracy can provide traders with a valuable edge over the competition. 



It is not surprising therefore that many researchers have attempted to use a variety of data analysis and machine learning techniques~\cite{hu2021DLsurvey} to extract meaningful patterns from market data whether attempting to determine the fair value of a stock (so-called \emph{fundamental analysis}~\cite{haque2013fundamental}) or predicting its future trajectory (so-called \emph{technical analysis} \cite{kumar2020technical}). For example, traditionally, stock returns prediction has been tackled using statistical techniques such as autoregressive integrated moving average models~\cite{ARIMA_ariyo2014}, but with recent advances in machine learning and deep learning, research applying advanced computational techniques to the problem of stock market prediction has become increasingly popular~\cite{Ozbayoglu2020_DL_Survey}. 

Indeed, the potential role for case-based reasoning (CBR) in financial domains was discussed early on in the CBR literature \cite{slade1991case} and there have been numerous attempts to apply case-based ideas to a variety of financial decision making and prediction tasks over the years \cite{kim2004toward,oh2007financial,li2009predicting,wang2016case,Chang2011,chun2020geometric} with varying degrees of success. However, one of the problems facing similarity-based methods concerns the challenge of developing a suitable similarity metric with which to assess the similarity of price-based time series. For example, conventional Euclidean distance and correlation based metrics have typically fallen short, leading some researchers to explore alternatives; see for example, Chun and Ko's \cite{chun2020geometric} shape-based distance metric.

In this paper we apply case-based reasoning techniques to stock selection based on the prediction of future returns, using only historical pricing data; in Section \ref{section:cases} we describe the basic case representation. The main contribution is the development of a novel hybrid similarity metric combining information about price deviations and trends into a single metric; see Section \ref{section:similarity_metrics}. Then, in Section \ref{section:trading} we present the results of a comprehensive offline evaluation of this approach by evaluating the returns produced by trading strategies using this approach, and in comparison to a variety of alternative benchmarks, to demonstrate significant benefits due to our approach across a range of suitable evaluation metrics.

\section{Related Work}
As a reuse-based problem solving method, guided by similarity \cite{aamodt1994case}, case-based reasoning is an appealing paradigm when it comes to a variety of decision problems in financial domains. Intuitively, the idea of basing current decisions on the outcomes of similar decisions that have been made in the past --- the core of CBR --- seems like an excellent fit in many financial settings. Even though historical patterns will not always prove to be a reliable guide to the future, markets are often driven by cyclical patterns and seasonal trends, which can be exploited to good effect. Indeed case-based reasoning has had a long history when it comes to tackling a range of important problems in financial domains, with applications spanning several distinct topics such as bond rating prediction~\cite{shin1999bond1, shin2001bond2}, bankruptcy~\cite{jo1997bankruptcy1, ahn2009bankruptcy2, alaka2018bankruptcy3}, financial risk assessment~\cite{kapdan2019risk}, real estate valuation~\cite{yeh2018realestate} and stock market prediction~\cite{chun2004mining, chun2005ensemble, chun2006regression, Chang2009, goswami2009candlestick, Chang2011, bedo2013similarity, Ince2014, chun2020geometric}. 

While recent work on the application of case-based reasoning to stock market prediction have been somewhat scarce \cite{chun2020geometric}, previous efforts have explored a variety of approaches in terms of their case representations and similarity metrics. Often cases are represented as (multivariate) time series \cite{chun2004mining, chun2005ensemble, chun2006regression} but sometimes more conventional feature-based approaches are used; \cite{Ince2014} selects twelve fundamental and technical indicators as predictor variables. In this paper, cases are represented by a simplified univariate time series using historical monthly returns.

When it comes to case similarity, the literature discusses a variety of options including conventional approaches such as 
 Euclidean, Manhattan and Gaussian distance metrics~\cite{chun2004mining, chun2005ensemble, chun2006regression}; ~\cite{Ince2014} proposes the use of genetic algorithms to determine the feature weights in a Euclidean distance metric. One of the problems with such metrics is that they fail to adequately account for the temporal nature of time-series data such as pricing data ~\cite{chun2020geometric}. This has motivated recent work by Chun and Ko~\cite{chun2020geometric}to develop a more geometrically inspired approach to time-series similarity. Their \emph{shape-based} similarity metric focuses on the patterns of rising and falling price-data, between two time series, rather than on the differences between prices at a given point in time. The work presented in this paper is similarly motivated and we too propose a new similarity metric as the centrepiece of our CBR approach to stock selection and returns prediction.
 


\section{From Prices to Cases}\label{section:cases}
The dataset used in this work spans the fifteen-year period from 01/01/2005 to 01/01/2021. Assets were selected from a range of international markets with the inclusion criterion being: (i) the availability pricing data spanning the period in question and (ii) their inclusion in the Nasdaq 100, EURO STOXX 50 or FTSE 100 indices. The resulting dataset was downloaded from Yahoo! Finance and contained 160 unique stock/asset tickers from six stock exchanges. 

The resulting raw data consisted of daily adjusted closing prices for each stock. When considering the problem of stock price prediction, a common approach in the literature has been next-day price prediction~\cite{yang2018explainable, long2020nextday}, with some considering even shorter time spans~\cite{alostad2015hourly, selvin2017minutes}. However, stock market returns are notoriously hard to predict, especially for shorter time spans due to the increased influence of market noise on price movements \cite{ivestopedia}. Thus we first convert the raw daily data into monthly price data with each $p$ indicating the price of a stock at the end of a given month. Then we transform the monthly pricing data into monthly returns data in order to extract a more reliable signal (see Equation \ref{eqn:to_returns}). 

\begin{equation}
\label{eqn:prices}
    prices(a_i) = \{p^{a_i}_1, ..., p^{a_i}_n\}
\end{equation}

\begin{equation}
\label{eqn:to_returns}
    r_{t}^{a_i} = \frac{p_{t}^{a_i}-p_{t-1}^{a_i}}{p_{t-1}^{a_i}}
\end{equation}
Accordingly, each case, for asset $a_i$ at time $t$ ($c(a_i,t)$) consists of a sequence of monthly returns for the previous twelve months (the \emph{problem description} part of the case) and a corresponding return for the single next month (the \emph{solution} part of the case); see Equation \ref{eqn:case_equation}. 
\begin{equation}\label{eqn:case_equation}
    c(a_i,t)=\{r_{t-12}^{a_i}, r_{t-11}^{a_i}, ..., r_{t-1}^{a_i}~|~  r_{t}^{a_i}\}
\end{equation}

Obviously, this case structure is a very simple one, purposely so. It has been chosen for two main reasons. First, it is a good fit for the type of similarity metric that we develop in the following section. Second, by simplifying our case structure in this way we can avoid the many additional factors that may complicate performance analysis and obscure the reason for a particular evaluation outcome, not to mention limiting the explainability of this approach. Indeed, we suggest that if we can generate good predictions using this simple case structure then it suggests an effective performance baseline and a platform for future enhancements.

This case structure was used to build a case base as follows. First, for reasons of computational efficiency, we limited our data to the period between January 2005 and December 2020 (180 months in total). Then, for each of the 160 companies/stocks in our dataset, we constructed cases during this period, with each case containing the past returns for the preceding 12 months and the return for the current (13th) month leading to 28,880 (180$\times$160) unique cases. Later we will discuss how this case base was used during our evaluation.


\section{Similarity in Financial Time Series}\label{section:similarity_metrics}
Similarity is obviously central to case-based reasoning but conventional similarity metrics such as Euclidean distance or cosine similarity tend not to fare well when it comes to assessing time-series similarity because they ignore the temporal relationship that exists between the different feature values, or monthly returns in this case. In this section, we propose a new similarity metric that emphasises two aspects of similarity that are important in a financial setting: (i) the correlation between time-series returns; and (ii) similar cumulative returns at the end of an investment period. In other words, given a target query case $q$, we wish to identify a set of similar cases whose monthly returns behave in a manner that is similar to the monthly returns of $q$ and whose cumulative return is similar to $q$'s cumulative return.

As an aside, at this stage it is worth highlighting a somewhat unusual and counter-intuitive feature of similarity assessment in a stock-trading setting. Namely, it is not only important to be able to identify a set of similar stocks, but also a set of dissimilar stocks that are expected to behave in opposition to the similar stocks. This is because, in a trading context, traders will often need to offset or hedge their positions in selected stocks by also trading in maximally dissimilar stocks; the idea being that under-performance in a selected (similar) stock can be offset by gains in a dissimilar stock, thereby allowing a trader to limit their overall risk. While we do not consider this aspect in more detail in this paper it is nevertheless an important consideration when selecting a suitable similarity metric and we will comment on this further below.

\subsection{The Problem with Correlation}
It is a commonly held belief, by investors, and even some academics, that a positive correlation between two stock-price time-series indicates that the stocks move in the same direction at the same time, while a large negative correlation indicates that the asset tends to move in opposite directions \cite{lhabitant2020correlationEDHEC}. In fact, a correlation-based metric, such as Pearson's, actually tends to measure the degree to which the returns deviate above or below their mean at the same time. This distinction is significant in the financial domain and will be highlighted below through an example.

\begin{figure}[!htb]
     \centering
     \begin{subfigure}[b]{0.48\textwidth}
         \centering
         \includegraphics[width=\textwidth]{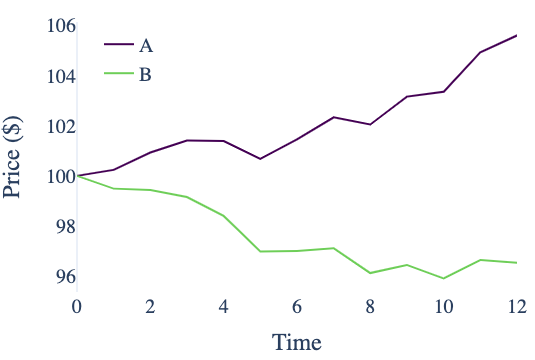}
         \caption{Sample Asset Paths}
         \label{fig:corr_demo1}
     \end{subfigure}
     \hfill
     \begin{subfigure}[b]{0.48\textwidth}
         \centering
         \includegraphics[width=\textwidth]{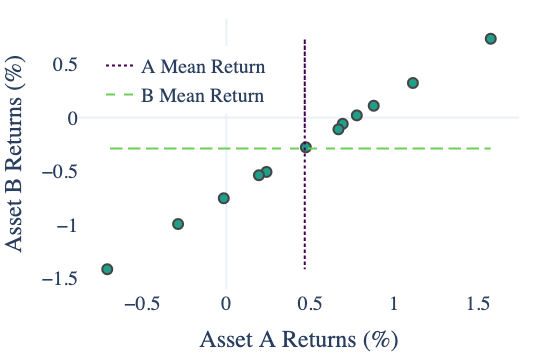}
         \caption{Returns - Asset A vs. B}
         \label{fig:corr_demo2}
     \end{subfigure}

        \caption{Correlation Example}
\end{figure}

Consider the price evolution of two hypothetical assets A and B in Figure~\ref{fig:corr_demo1}, but note that correlation is calculated based on \emph{returns} (differences in prices) rather than prices, in order to discount the underlying trends that would otherwise overly influence the correlation. In other words, the correlation between A and B is based on the sequence and magnitude of their price changes rather than the actual prices themselves. The point is that an investment in asset A performed well over the period, with a consistent positive return, while an investment in asset B lost money. Despite this, a traditional correlation metric such as Pearson's correlation coefficient (see Equation \ref{eqn:pearson}) determines that they are perfectly positively correlated; Pearson's returns a near perfect correlation value of 0.99 in this case. This is illustrated in Figure \ref{fig:corr_demo2} which shows each pair of monthly returns from Figure \ref{fig:corr_demo1} as a set of points with a clear linear correlation.


\begin{equation}\label{eqn:pearson}
\displaystyle
    \rho(x,y) =\frac{\sum ^n _{i=1}(x_i - \bar{x})(y_i - \bar{y})}{\sqrt{\sum ^n _{i=1}(x_i - \bar{x})^2 \sum ^n _{i=1}(y_i - \bar{y})^2}}
\end{equation}

\subsection{An Adjusted Correlation Metric}\label{section:adjustd_correlation}

This correlation phenomenon is particularly problematic in a financial setting and it is a known problem with conventional correlation, such as Pearson's correlation coefficient \cite{lhabitant2020correlationEDHEC} but few practical solutions have been proposed. One actionable diagnosis of the problem is that it occurs because individual monthly returns are assessed relative to \emph{mean} returns ($\bar{x}$ and $\bar{y}$ in Equation \ref{eqn:pearson})~\cite{quantdare}. Asset A has an overall positive mean return, compared with an overall negative mean return for Asset B, leading to the positive correlation. Thus, a straightforward solution is to derive a modified correlation by simply eliminating the dependency on the means and instead shifting the point of reference to zero. This modification is shown in Equation \ref{eqn:new_correlation}

\begin{equation}\label{eqn:new_correlation}
\displaystyle
    \tau(x,y) =\frac{\sum ^n _{i=1}x_i \cdot y_i}{\sqrt{\sum ^n _{i=1}x_i^2 \cdot \sum ^n _{i=1}y_i^2}}
\end{equation}

In what follows, we will refer to this as the \emph{adjusted} correlation metric. Similar to Pearson's correlation metric, this adjusted metric returns values in the interval [-1, +1] and in the case of the data shown in Figure \ref{fig:corr_demo1}, this adjusted metric returns a value of 0.425.

\subsection{A Novel Similarity Metric for Returns-Based Time-Series}
We mentioned earlier that it is desirable for our metric to measure similarity in terms of the tendency for a pair of stock cases to deliver similar returns at similar times -- the adjusted correlation metric provides for this -- but also to ensure that their cumulative returns are similar. To address the latter requirement we propose using Equation \ref{eqn:cumprod} which calculates the relative difference between two cases, $c(a_i,t)$ and $c(a_j,s)$, based on the product of their monthly returns; this product of monthly returns is mathematically equivalent to the relative difference between the start and end price of each stock over their 12 month periods, but since cases are represented using returns data, rather than price data, we calculate the cumulative return in this way.

\begin{equation}\label{eqn:cumprod}
    \displaystyle e(c_{a_i,t},c_{a_j,s})=\sqrt{\left(\prod_{\hat{t}=t-1}^{t-12}(1+r_{\hat{t}}^{a_i}) - \prod_{\hat{s}=s-1}^{s-12}(1+r_{\hat{s}}^{a_j})\right)^2}
\end{equation}

Then, we present our overall similarity metric as Equation \ref{eqn:new_metric}, which calculates the cumulative returns and adjusted correlation metric; note that the cumulative returns metric has been incorporated in Equation \ref{eqn:new_metric} in such a way that it serves as a true similarity metric, rather than a distance metric.

\begin{equation}\label{eqn:new_metric}
    sim(c_{a_i,t},c_{a_j,s}) =  \frac{w}{1+e(c_{a_i,t},c_{a_j,s})} + (1-w)\cdot\tau(c_{a_i,t},c_{a_j,s})
\end{equation}

\begin{figure}
     \centering
     \begin{subfigure}[b]{0.48\textwidth}
         \centering
         \includegraphics[width=\textwidth]{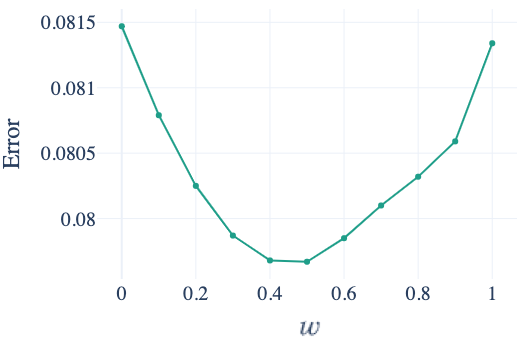}
         \caption{Error vs. Weight}
         \label{fig:weight_plot}
     \end{subfigure}
     \hfill
     \begin{subfigure}[b]{0.48\textwidth}
         \centering
         \includegraphics[width=\textwidth]{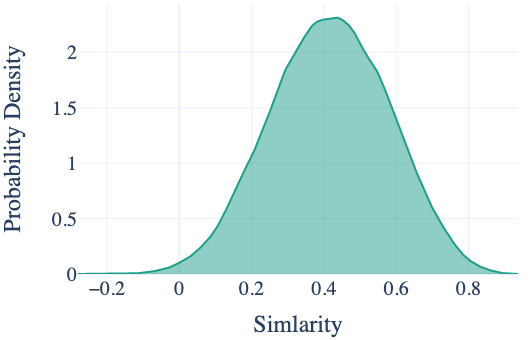}
         \caption{Overall Distribution}
         \label{fig:overall_distribution}
     \end{subfigure}

        \caption{Analysis of Proposed Metric}
        \label{fig:weight_overall}
\end{figure}

Obviously, the relative importance of the cumulative returns and correlation components can be adjusted by varying $w$; when $w=0$ the similarity equation is based solely on the adjusted correlation metric and when $w=1$ it resorts to euclidean distance between cumulative returns only. In order to evaluate the impact of $w$ on similarity, using each case in our case base as a \emph{query} we calculate the top-20 most similar cases using the above metric with different values of $w$ ($0\leq w \leq1$) and then compare the next-month returns for the similar cases to the actual next-month return for the corresponding query cases. The absolute relative difference between the return of the similar cases and the query case serves as an error score and the mean error score by $w$ is shown in Figure \ref{fig:weight_plot}. We can see that the optimal error occurs for values of $w$ between 0.4 and 0.5 and for the remainder of this study we use $w=0.5$; obviously, this optimal weight may be sensitive to different case bases and case structures. Figure \ref{fig:overall_distribution} shows a histogram of the similarity values obtained during this analysis; the results suggest that the metric behaves as expected as a similarity metric.


\subsection{Most and Least Similar Cases}

\begin{figure}
     \centering
     \begin{subfigure}[b]{\textwidth}
         \centering
         \includegraphics[width=\textwidth]{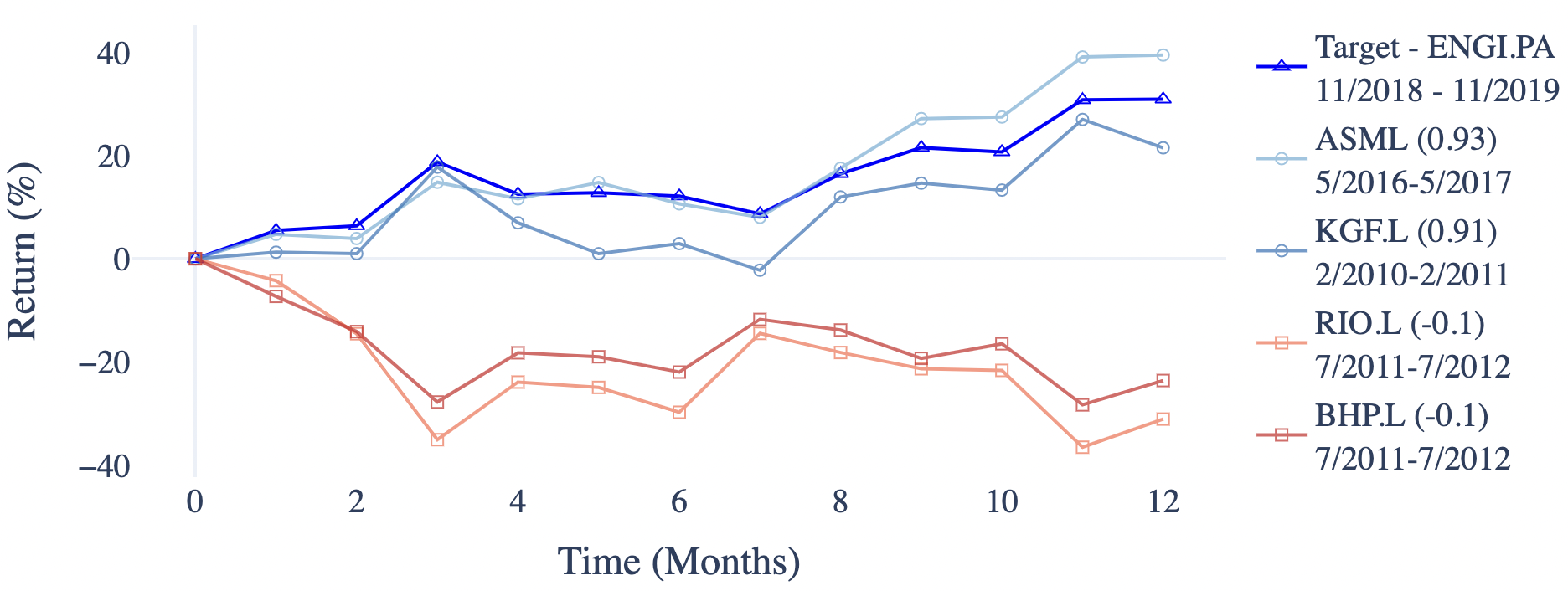}
         \caption{The query case represents the 12 month price evolution of Engie SA.}
         \label{fig:sim1}
     \end{subfigure}
     \begin{subfigure}[b]{\textwidth}
         \centering
         \includegraphics[width=\textwidth]{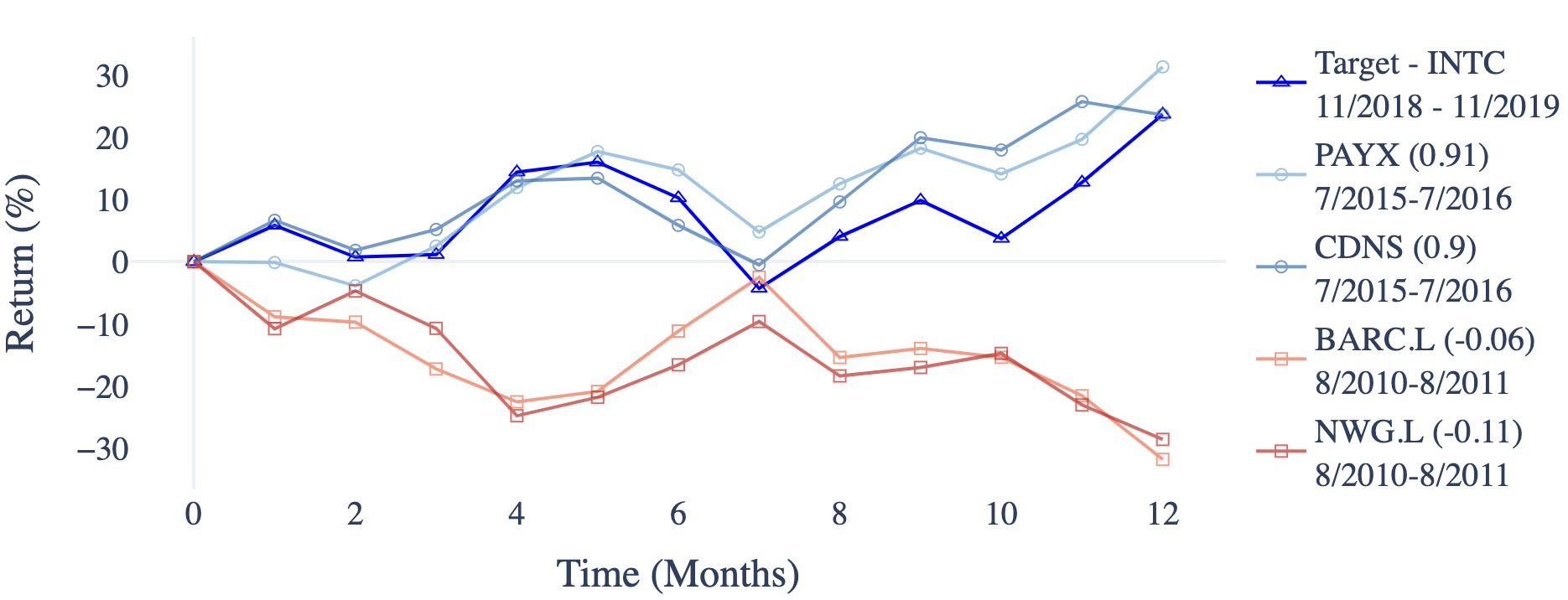}
         \caption{The query case represents the 12 month price evolution of Intel Corporation.}
         \label{fig:sim2}
     \end{subfigure}
     \begin{subfigure}[b]{\textwidth}
         \centering
         \includegraphics[width=\textwidth]{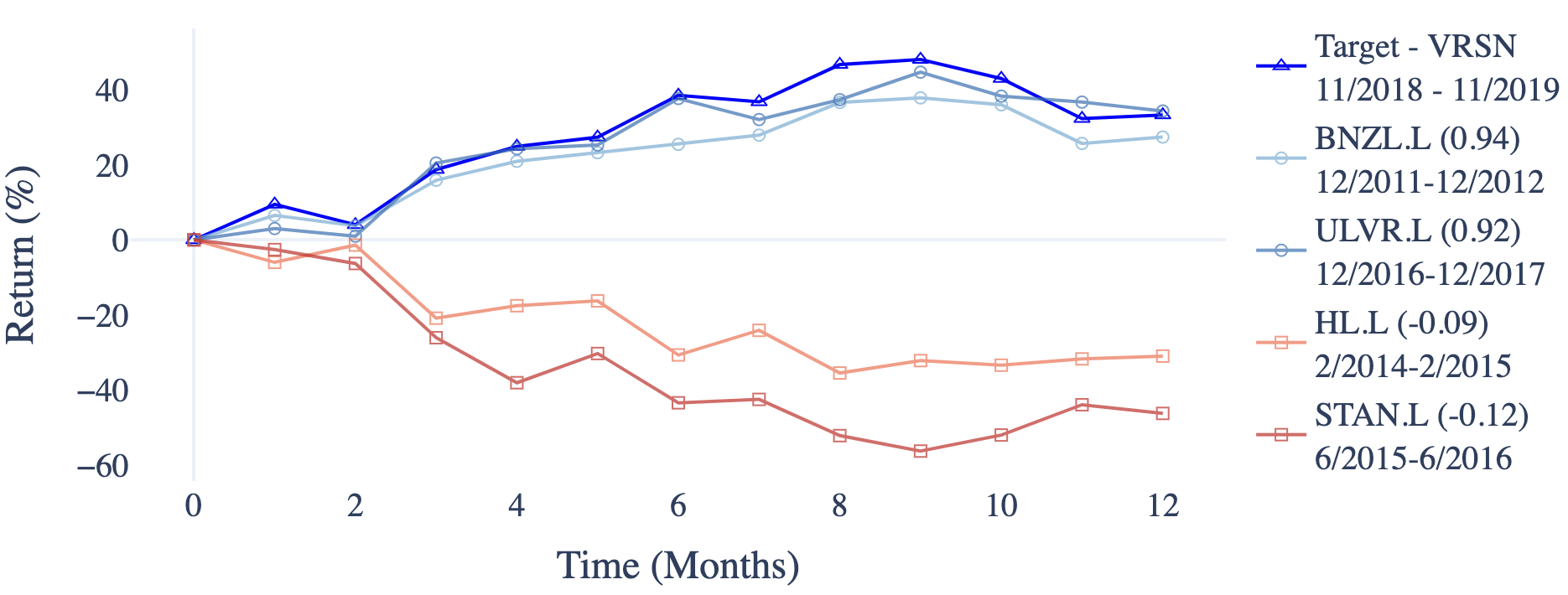}
         \caption{The query case represents the 12 month price evolution of Verisign.}
         \label{fig:sim3}
     \end{subfigure}

        \caption{Example of the two most and least similar cases for randomly chosen query cases in the time period 11/2018 -- 11/2019. The two most similar cases are plotted in light blue with circle symbols while the two least similar cases are plotted in red with square symbols. The asset ticker, similarity and time period for each similar case is given in the legend.}
        \label{fig:sim_cases_example_overall}
\end{figure}

Visualising the most and least similar cases for a randomly selected query case is a simple way of verifying the efficacy of the proposed similarity metric and we do this for three separate examples in Figure \ref{fig:sim_cases_example_overall}. Taking Figure \ref{fig:sim1} as an example, it illustrates the two most and least similar cases for the target query defined by asset Engie SA over the time period 11/2018 to 11/2019. It is seen that the price evolution of the two most similar cases track that of the query case very closely. Not only do the high similarity cases exhibit similar cumulative returns (end up very close), but they also tend to rise and fall at the same points in time. Conversely, the two least similar cases almost mirror the query case over the x-axis. Firstly, their cumulative returns are highly negative in contrast with the strong positive cumulative return of the query case. Secondly, the deviations at each point in time tend to be opposite in sign but similar in magnitude to that of the query case, as we would hope. This is particularly evident at time 3, for example, where the query has a strong positive return while both low similarity cases have very large negative returns for that month.

In conventional approaches, the tendency is to focus on either the deviations at each individual point in time \emph{or} the overall trend. For example, both Pearson's correlation and Chun's~\cite{chun2020geometric} recent geometrical similarity metric focus solely on the rises and falls at each individual time period but disregard the overall trend. The novel formulation proposed in Equation \ref{eqn:new_metric} allows us to capture both of these components simultaneously as the examples in Figure \ref{fig:sim_cases_example_overall} illustrate.

Interestingly, the most and least similar cases in Figure \ref{fig:sim1}, for example, are all from different markets to the target. The target, Engie SA (ENGI.PA) is a French company listed on the Euronext stock exchange in Paris while its most similar case ASML Holdings (ASML) is traded on the NASDAQ exchange in the USA. The second most similar case and the two least similar cases are all listed on the London Stock Exchange. Additionally, we note that the most and least similar cases occur up to eight years prior to the target case with both low similarity cases coming from the same 12-month period.

\section{Evaluation}\label{section:trading}
So far we have described a case representation for encoding the relationship between the previous 12 months of returns for a given stock and the next month of returns, and we have presented a novel similarity metric, which we believe can provide a better sense of similarity in this task domain. In this section, we will describe the results of an evaluation to compare the performance of this new metric to a variety of alternatives in an investment setting\footnote{The relevant code can be found at \href{https://github.com/rian-dolphin/ICCBR2021-Financial-TS-Similarity}{https://github.com/rian-dolphin/ICCBR2021-Financial-TS-Similarity}}. In fact, we will conduct two related evaluations: (1) predicting next-month returns; and (2) using predicted next-month return to inform stock selection as part of an extended investment strategy. In the former we will compare our proposed metric to a variety of alternatives in terms of their ability to accurately predict next-month returns. In the latter we will use these predictions to select the top-5 stocks with the highest predicted returns each month, over a 172 month period, to compare different strategies in terms of the compounded, cumulative investment gains.

In both evaluations we compare the results obtained using the following different similarity variations:
\begin{enumerate}
    \item $ProposedAdjusted$, the main similarity metric proposed in this paper which combines adjusted correlation and the Euclidean distance between cumulative returns.    
    \item $ProposedPearson$, the similarity metric proposed in this paper but using Pearson correlation instead of adjusted correlation.
    \item $PearsonOnly$, a conventional time-series similarity measure using Pearson's correlation metric.
    \item $Shape$, the authors' version of the geometric similarity metric described by \cite{chun2020geometric}.
    \item $AdjustedOnly$, the adjusted form of Pearson's correlation from Equation \ref{eqn:new_correlation} and used in $Proposed$.
    \item $CumulativeOnly$, the Euclidean distance between cumulative returns metric from Equation \ref{eqn:cumprod}.
\end{enumerate}

\subsection{Predicting Monthly Returns}

\begin{figure}[!tb]
    \centering
    \includegraphics[width=1\textwidth]{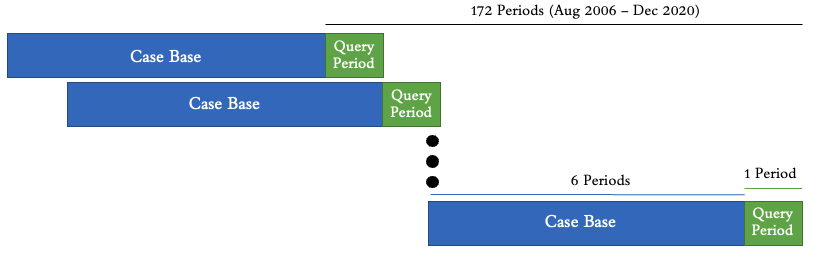}
    \caption{Rolling Window Layout}\label{fig:windows}
\end{figure}

In this part of the evaluation, the goal is to predict the next-month returns for a stock based on its previous 12 months of returns. Due to the temporal nature of the data, care must be taken to ensure that only cases that refer to periods prior to the query case period are considered during retrieval; thus if we wish to predict the next-month return for March 2019 then we can only draw on cases whose next-month returns occur prior to March 2019. As a result, a simple leave-one-out strategy cannot be directly implemented, and so, we employ a rolling window approach inspired by~\cite{chan2016rollingwindow, bao2017rollingwindow}, but tailored to a CBR framework. This approach, illustrated in Figure \ref{fig:windows}, allows us to utilise as many query cases as possible in our evaluation but has the effect that the case base depends on the query case. In particular, the case base is defined to contain all cases from the previous six time periods, with six being chosen due to computational limitations. Equation \ref{eqn:base} formalises the case base, $\mathcal{C}(c(a_i,t))$, for a general query case $c(a_i,t)$.


\begin{align}\label{eqn:base}
  \mathcal{C}\Big(c(a_i,t)\Big) &= \left\{ c(a_i,t-j)\ \middle\vert \begin{array}{l}
    i\in\{1,2,...,160\} \\
    j\in\{1,2,...,6\}
  \end{array}\right\}
\end{align}

For each query case $q$, the task is to predict $q$'s next-month returns based on a similarity-weighted mean of the next-month returns for the $k$ most similar cases to $q$. Each prediction is compared to the actual next-month returns for $q$ via an absolute difference, giving us an error measurement. We repeat this for different values of $k$ from 1 to 50.


\begin{table}[!htb]
\centering
\caption{Mean errors in next-month returns for varied $k$ and similarity metrics. $H_0$ refers to the result of a Tukey HSD test with null hypothesis that the mean of the $ProposedAdjusted$ metric is not significantly different from the given baseline metric with $k=10$. Rejection of the null hypothesis at $\alpha=0.01$ is indicated by \cmark.}
\begin{tabular}{lcccccc}
\toprule
{} &       \hspace{0.5cm}1\hspace{0.5cm}  &           \hspace{0.5cm}5\hspace{0.5cm}  &       \hspace{0.5cm}10\hspace{0.5cm} &       \hspace{0.5cm}25\hspace{0.5cm} &       \hspace{0.5cm}50\hspace{0.5cm} & \hspace{0.2cm}$H_0$\hspace{0.2cm} \\
\midrule
$ProposedAdjusted$     &  \textbf{0.0887}  &  \textbf{0.0882} &  \textbf{0.0883} &  \textbf{0.0882} &  \textbf{0.0883} & - \\
$ProposedPearson$ &  0.0893 &   0.0885 &  0.0884 &  0.0883 &  0.0884 & \xmark \\
$Shape$           &  0.0888 &    0.0889 &  0.0889 &  0.0891 &  0.0893 & \cmark \\
$PearsonOnly$            &  0.0910 &    0.0909 &  0.0909 &  0.0909 &  0.0908 & \cmark \\
$CumulativeOnly$           &  0.0911 &    0.0908 &  0.0907 &  0.0904 &  0.0903 & \cmark \\
$AdjustedOnly$   &  0.0911 &    0.0908 &  0.0908 &  0.0907 &  0.0905 & \cmark \\
\bottomrule
\end{tabular}\label{tab:error_table}
\end{table}

The mean error results presented in Table \ref{tab:error_table} show how the proposed metric tends to produce predictions with lower error rates than all of the other variations considered. Although the differences are small, it must be remembered that this reflects the errors associated with a single monthly prediction and obviously these error have the potential to compound and accumulate if their corresponding metrics are used to inform an extended trading strategy over time. We will return to this in the section that follows.

Post-hoc Tukey HSD tests confirm that there are significant differences between the pairs of techniques shown in Table \ref{tab:error_table}. Although the $ProposedAdjusted$ technique shows significant improvement with respect to the $PearsonOnly$, $AdjustedOnly$, $Shape$, and $CumulativeOnly$ metrics at $k=10$ it is not significantly better than $ProposedPearson$, at least in terms of the error associated with a single monthly returns prediction. It is worth noting, however, that $ProposedPearson$ is the only other strategy, in addition to $ProposedAdjusted$, which uses the novel formulation proposed in Equation \ref{eqn:new_metric}.

\subsection{Comparing Trading Strategies}

As mentioned above, the previous experiment focused on a single next-month returns prediction, but in practice trading performance is measured over an extended period of time based on the cumulative returns obtained during many buy-sell cycles. In order to evaluate this, in this section we consider a simple trading scenario in which a trader begins with a \$1,000 float and invests this uniformly in the stocks with the top-5 highest predicted returns each month, selling these stocks at the end of the month, and rolling-up any profits/losses into their next month investment. This continues for a period of 172 months, as outlined in Figure \ref{fig:windows}, and the cumulative gains are calculated at the end of this period. We use the six different similarity metrics, as before, to generate the monthly returns predictions, and various values of $k$ are used, also as before. 

Since we are simulating buying the top-5 assets in terms of the highest predicted return each month, the strategies will execute 860 (5 trades $\times$ 172 months) buy orders in total over the course of the experiment. Though a sizable number of trades, running the simulation only once would mean the evaluation has the potential to be influenced by a small number of `lucky' trades. To prevent this, we ran the simulation one hundred times, each time randomly removing 20\% of the assets from the dataset.

\begin{table}[!tb]
\centering
\caption{Results of a trading simulation spanning mid 2006 - end 2020 with $k=10$ and initial capital of \$1000. $H_0$ refers to the result of a Tukey HSD test with null hypothesis that the mean annualised return of the $ProposedAdjusted$ metric is not significantly different from the given baseline metric with $k=10$. Rejection of the null hypothesis at $\alpha=0.01$ is indicated by \cmark.}
\begin{tabular}{lcccccc}
\toprule
{} &       \hspace{0.2cm}\thead{Accumulated\\Value}\hspace{0.2cm}  &  \hspace{0.2cm}\thead{Annualised\\Return}\hspace{0.2cm}  &           \hspace{0.2cm}\thead{Annualised\\Volatility}\hspace{0.2cm}   & \hspace{0.2cm}$H_0$\hspace{0.2cm} \\
\midrule
$ProposedAdjusted$     &  \textbf{\$8,305.23}  &    \textbf{15.9\%} &  22.3\%   & - \\
$ProposedPearson$ &  \$6,551.87 &    14.0\% &  22.2\% & \cmark \\
$Shape$           &  \$7,797.25 &   15.4\% &  \textbf{19.7\%} & \xmark \\
$PearsonOnly$     &  \$6,233.49 &  13.6\% &  21.3\%  & \cmark \\
$CumulativeOnly$  &  \$6,527.48 &    14.0\% &  21.6\%  & \cmark \\
$AdjustedOnly$    &  \$7,883.33 &    15.5\% &  21.1\%   & \xmark \\
\bottomrule
\end{tabular}\label{tab:trading_table}
\end{table}


The results are presented in Table \ref{tab:trading_table} for each strategy with $k=10$. Under the $ProposedAdjusted$ strategy the initial float of \$1000 accumulates to \$8305.23, corresponding to an annualised return of 15.9\%, the highest of all the strategies. A Tukey test indicates the mean return for $ProposedAdjusted$ is significantly higher than that of $ProposedPearson$, $PearsonOnly$ and $CumulativeOnly$. Though a higher mean return is seen, the Tukey test does not confirm statistical significance over $Shape$ and $AdjustedOnly$ at $\alpha=0.01$. 

It is notable too that the $AdjustedOnly$ strategy beats the $PearsonOnly$ strategy (significantly at $\alpha=0.01$) indicating that the modified correlation metric described in Section \ref{section:adjustd_correlation} is outperforming the more conventional Pearson correlation metric when applied in a trading simulation. In fact, Pearson correlation alone performs worse than all other strategies. Chun's~\cite{chun2020geometric} more recent geometrical similarity metric ($Shape$) performs well in this trading evaluation, producing annualised returns that are better than most of the other strategies, although not the proposed strategy. Its performance is very similar to that of the $AdjustedOnly$ similarity metric which is unsurprising since the adjusted correlation can, in some sense, be thought of as a continuous version of the geometric metric as both use 0 as a reference point. 

As predicted, although the individual monthly gains in prediction accuracy are small, when compounded as part of a selective investment strategy, then even modest gains can accumulate to offer significant differences in annualised returns. Obviously, this experiment represents a very simplified trading scenario that is limited by factors such as the number of stocks selected for investment each month and how current funds are shared among the selected stocks. In reality, one would expect more sophisticated trading strategies to be used, which vary the number of stocks selected each month and how funds are divided up for the purpose of investment. It may be prudent to include other indicators to aid the trade selection process and it may also be appropriate to consider inter-stock similarity when selecting stocks to provide some level of hedging/diversification as part of an investment strategy. All of these factors will further influence the returns obtained and none have been considered in this initial evaluation.

\section{Conclusion and Future Work}

This work has focused on the problem of measuring similarity between financial time series with a particular focus on stock market pricing and returns data. Our proposed metric combines an adjusted correlation coefficient with a Euclidean metric to simultaneously identify similarity from two angles which, to the best of our knowledge, has not been explored before. 

In addition, we have applied our novel similarity metric to the problem of predicting stock market returns and using this to inform a trading strategy. Although this is no doubt a challenging problem, it is motivated by the knowledge that even modest returns and improvements can prove to be extremely useful in the high-stakes world of finance. 

We have described a straightforward approach to using ideas from case-based reasoning for this task, including a simple returns-based case representation and a novel approach to measuring the similarity between stock time-series. The results of an initial evaluation demonstrate strong results in terms of returns-based prediction accuracy which in turn lead to significant benefits in terms of annualised returns when used as part of a stock trading strategy. Moreover, the results reported for our novel similarity-metric are superior to those for a variety of alternative including conventional and state-of-the-art baselines. 



There is substantial scope for future work with the approach described in this paper. The trading strategy used during the evaluation is likely too simple to be useful in practice and can be enhanced in a number of ways to more reliably evaluate the benefits of the new similarity metric. Comparing our results to state-of-the-art non-CBR baselines such as long short-term memory (LSTM) networks as well as testing varied case lengths are other planned areas of future work. Moreover, modern portfolio theory is based heavily on the use of Pearson correlation to ensure diversification and there is an obvious opportunity to evaluate our revised similarity metric and the adjusted correlation coefficient in the portfolio optimisation domain.

\textbf{Acknowledgements.} This publication has emanated from research conducted with the financial support of Science Foundation Ireland under Grant number 18/CRT/6183.

%
%
%
%
\bibliographystyle{splncs04}
\bibliography{main.bib}

\end{document}